# Analysis of Email Fraud Detection Using WEKA Tool

Author:Tarushi Sharma , M-Tech(Information Technology),

*CGC Landran Mohali, Punjab,India,*

Co-Author:Mrs.Amanpreet Kaur (Assistant Professor),

*CGC Landran Mohali, Punjab,India*

*Abstract*—Data mining is also being useful to give solutions for invasion finding and auditing. While data mining has several applications in protection, there are also serious privacy fears. Because of email mining, even inexperienced users can connect data and make responsive associations. Therefore we must to implement the privacy of persons while working on practical data mining.

Using K-mean clustering algorithm and weka tool we implemented the technique of Email-mining. The WEKA tool calls the .eml file format into text converter and then processed the whole data into preprocessor output in form of .csv file format. The preprocessor output shows the graphical results of the processed email data. The goal of this implementation is to detect or filter the email addresses from which we get maximum emails.

*Index Terms*— Data mining, Email mining, Weka tool, K-mean clustering algorithm, Preprocessor

## I. INTRODUCTION

Every day E-mail users receive hundreds of spam messages with a new content, from new addresses which are automatically generated by robot software. To filter spam with traditional methods as black-white lists (domains, IP addresses, mailing addresses) is almost impossible. Application of text mining methods to an E-mail can raise efficiency of a filtration of spam. Also classifying spam messages will be possible to establish thematic dependence from geographical [1].

This paper focuses on the work done to classify the textual spam E-mails using data mining techniques. Our purpose is not only to filter messages into spam and not spam, but still to divide spam messages into thematically similar groups and to analyze them, in order to define the social networks of spammers [2].

In this paper we proposed a dynamic clustering of data with simple k-means algorithm. The algorithm takes number of clusters (K) as the input from the user and the user has to mention whether the number of clusters is fixed or not. If the number of clusters fixed then it works same as K-means algorithm.

## II. METHODOLOGY

Initially historical method of research will be used to collect the relevant data through authentic literature, books journals etc, in this research I will study a past published research thesis related to Email Mining [3].

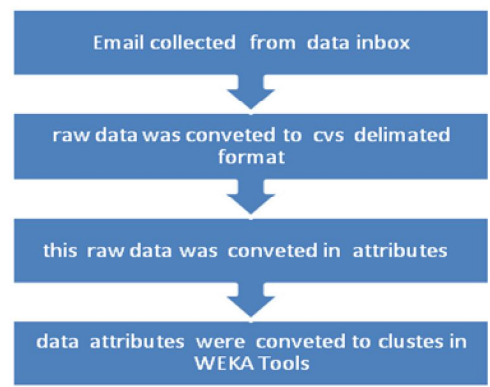

**Fig.1: Data analysis process structure**

After that evaluative and analytical method will be used through structured questionnaires of two patterns:

A. Primary Data: Information based on the results of analyzing data of logged emails and the research scholar will form the primary data.
B. Secondary data: Information based on the literature, books, journals will form the secondary data.
C. Survey method will be used for the study.
D. Tool and techniques for analysis would be Reliability and validity test.

Some of the data which was extracted as a single main from incoming mailbox different organizations or and personal mail boxes for sample I have collected data available for an organization in outlook express data formatted this data was converted to simple delaminated cvs format for easy data conversion [4].

### III. PROPOSED WORK

#### A. K-mean Algorithm

The k-means algorithm is an evolutionary algorithm that gains its name from its method of operation. The algorithm clusters observations into k groups, where k is provided as an input parameter [5, 11].

- **Input:**
  - k: the number of clusters.
  - D: a data set containing n objects.
- **Output:** A set of k clusters.
- **Method:**

1. Arbitrarily choose k objects from D as the initial cluster centres.
2. Repeat.
3. Re-assign each object to the cluster to which the object is most similar using Eq. 1, based on the mean value of the objects in the cluster.
4. Update the cluster means, i.e. calculate the mean value of the objects for each cluster.
5. Until no change.

#### B. WEKA

The workflow of WEKA would be as follows [6]:
- **Data → Pre-processing → Data Mining → Knowledge**
- The supported data formats are **ARFF, CSV, C4.5 and binary**. Alternatively you could also import from URL or an SQL database.
- After loading the data, preprocessing filters could be used for **adding/removing attributes, discretization, Sampling, randomizing** etc.

#### C. OutLook Express

Microsoft Outlook and Microsoft Exchange use a proprietary email attachment format called Transport Neutral Encapsulation Format (TNEF) to handle formatting and other features specific to Outlook such as meeting requests [7].

An open-source project called UnDBX was also created, which seems to be successful in recovering corrupt databases. Microsoft has also released documentation which may be able to correct some non-severe problems and restore access to email messages, without resorting to third-party solutions.

However, with the latest updates applied, Outlook Express now makes backup copies of DBX files prior to compaction. They are stored in the Recycle Bin. If an error occurs during compaction and messages are lost, the DBX files can be copied from the recycle bin [8, 14].

Opening or previewing the email could cause code to run without the user's knowledge or consent. Outlook Express does not correctly handle MIME, and will not display the body of signed messages inline. Users get a filled email and one attachment (one of the message text and one of the signatures) and therefore need to open an attachment to see the email [9].

## IV. RESULTS AND DISCUSSION

Process of Email mining through weka works as given below with results:

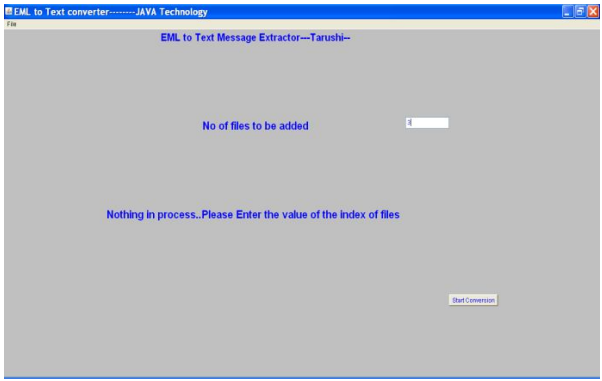

**Fig.2: Weka explorer EML to text converter**

The eml to text converter is the first dialog box of the implementation work which consist that how many emails are to be refine [9].

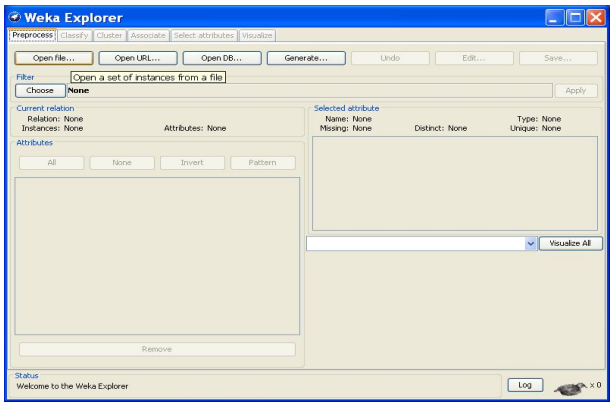

**Fig.3: Weka explorer showing Cluster output**

Open dialog box appeared to select the extract emails in form of .csv form which are located in Final directory. The WEKA tool calls the .eml file format into text converter and then processed the whole data into preprocessor output in form of .csv file format [10] . The preprocessor output shows the graphical results of the processed email data.

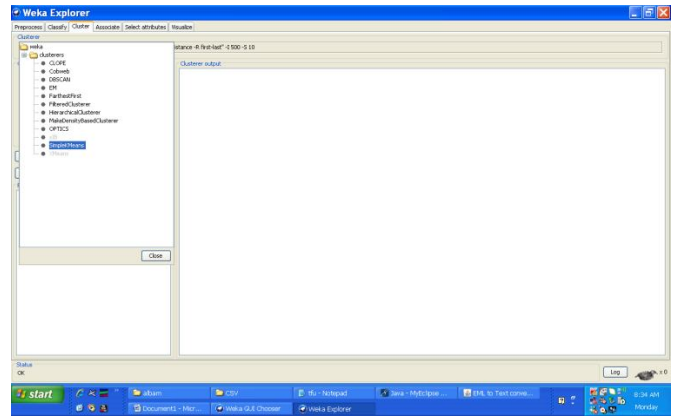

**Fig.4 Weka explorer to select K-mean algorithm**

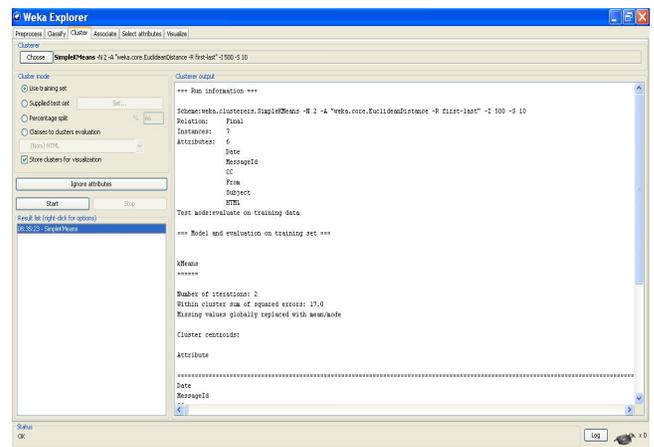

**Fig.5: Weka explorer showing Cluster output**

Cluster output shows:

| Parameters | Output |
|---|---|
| Instances | 7-number of live matches |
| Attributes | 6-Selected 6 attributes are Date, MessageId, CC, From, Subject, HTML |
| No. of Iterations | 2 |

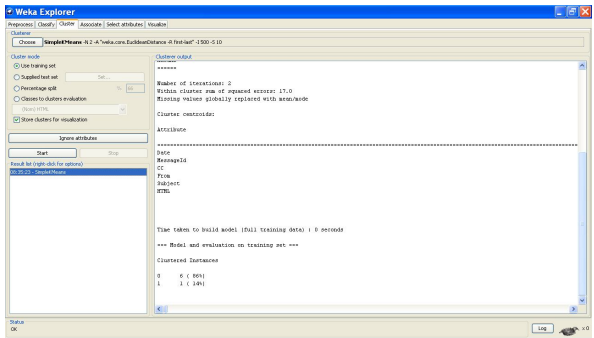

Fig.6: Weka explorer showing Cluster output

**Cluster output shows:**

| Parameters | Output |
|---|---|
| Instances | 6 (86%), means 6 out of 7 instances are identical |
| Instances | 1(14%) –means 1 out of 7 instance |
| No. of Iterations | 2 |

The graphical results shows the count and percentage values of instances on the basis of attributes selected. Attributes are the objects of email we select to compare [13]. The percentage values show the percentage amount of instances out of total. These instances shows the frequency of email data which helps in detecting the email locations that are sending maximum number of spam mails or unwanted mails. These are used to do email mining and filtering of emails [12].

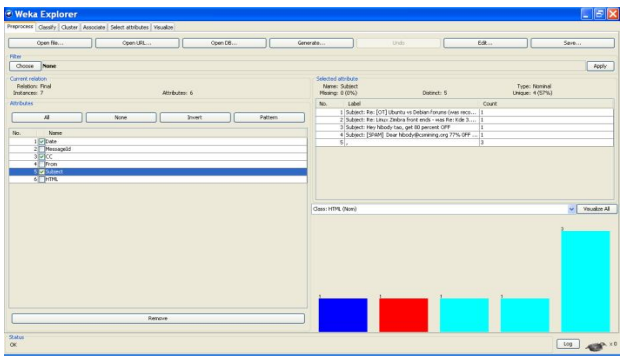

Fig.7: Weka explorer showing preprocessor output

**Cluster output shows:**

| Parameters | Output |
|---|---|
| Mails | 3 |
| Attributes | 3 out of 7 |
| Instances | 4 different and 3 identical |

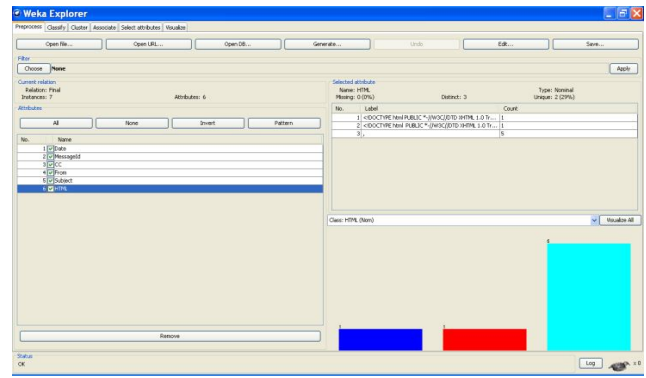

Figure 4.9: Weka explorer showing Preprocessor output

**Cluster output shows:**

| Parameters | Output |
|---|---|
| Mails | 3 |
| Attributes | 6 out of 7 |
| Instances | 2 different and 5 identical |

In second case we selected all 6 attributes and then processed the data of three mails. It resulted into 2 different instances and 5 similar instances of ",". And all those are shown in graphs of different colors.

## V. CONCLUSION AND FUTURE WORK

Coming to conclusion the main theme of this work is to detect or filter the emails to analyze the particular email address from which the number of receiving emails is maximized. A new idea that seems promising is Semantic Email has been proposed. The technique used is resultant to more secured feature to detect spam mails and their source address. The software implemented could be used to detect those methods and integrate them into useful and accurate email-mining which will let people take back control of their mailboxes.

Future recommendations have multiple choices likewise the present work is done for single email-id and the data processed is done for single email-id. But the work could be done by taking multiple email-ids. Moreover this filter evaluation technique is processed by taking just six attributes. But we can take much more number of attributes to improve the filtered results.

The efficiency of text converter which converts the outlook express file format into WEKA tool file formats. As in the present work the eclipse code generated perform the text conversion from .eml file format to .csv file format. But in further recommendations the file conversion could be perform for multiple file formats of WEKA tool like .arss. In this work we have chosen K-mean algorithm to process the clusters of data in WEKA tool. But in future work we can use more efficient existing or proposed algorithm to process the email data.

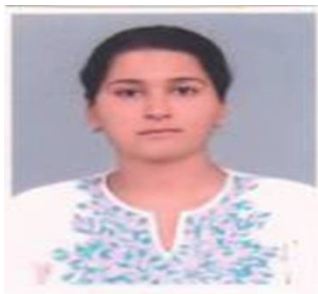

I am Tarushi Sharma persuing M-Tech from CGC Landran,Mohali(2011-13) in Information Technology.I did my B-Tech from CCS University Campus,Meerut(2007-11) in Information Technology.